\documentstyle[twocolumn,aps,prl,psfig]{revtex}
\begin{document}
\title{MODULATIONAL ESTIMATE FOR FERMI-PASTA-ULAM CHAIN LYAPUNOV
EXPONENTS}
\author{Thierry Dauxois\thanks{on leave from:
Laboratoire de Physique, URA-CNRS 1325,
ENS Lyon, 46 All\'{e}e d'Italie, 69364 Lyon C\'{e}dex 07, France}, 
Stefano Ruffo\thanks{INFN and INFM, Firenze (Italy)} 
and Alessandro Torcini\thanks{INFM, Firenze (Italy)}}
\address{Dipartimento di Energetica "S. Stecco", Universit\`a di Firenze,
via S. Marta, 3 , I-50139 Firenze (Italy)}
\date{\today}
\maketitle
\begin{abstract}
In the framework of the Fermi-Pasta-Ulam (FPU) model, we show
a simple method to give an accurate analytical estimation of
the maximal Lyapunov exponent at high energy density.
The method is based on the computation
of the mean value of the modulational instability
growth rates associated to unstable modes.
Moreover, we show that the strong stochasticity threshold
found in the $\beta$-FPU  system is closely related to a
transition in tangent space: the
Lyapunov eigenvector being more localized in space at high energy.
\end{abstract}
\pacs{PACSnumbers: 05.45.+b,03.20.+i,63.10.+a,03.40K}

A large number of theoretical and numerical studies have been devoted
to the characterization of chaotic high-dimensional systems,
but in spite of these efforts several fundamental items are not yet fully 
understood.
In particular, the relation between Lyapunov analysis
and other phase space properties like diffusion of orbits, relaxation 
to equilibrium states, spatial development of instability remains to 
be clarified~\cite{lichtlieb}.
This Rapid Communication, besides presenting an estimate of the largest Lyapunov exponent, 
is a contribution to clarify this relationship in the context
of the Fermi-Pasta-Ulam (FPU) model~\cite{ford}.
This system was not only
the starting point of the (re)discovery of the soliton~\cite{zabuskykruskal},
but initiated also several studies of the mixed
chaotic/ordered phase space structure based on
resonance overlap criteria~\cite{izrailev},
KAM theorem and Nekhoroshev stability estimates~\cite{galgani}.
Statistical mechanics was  also tested 
on this equation~\cite{bochieri,livi}
and the results
showed that ergodicity is not an obvious 
consequence of the non existence of
analytical first integrals of motion.

There are very few examples of analytical calculations
of the largest Lyapunov exponent $\lambda_1$
in the chaotic component corresponding to energy equipartition
in high-dimensional systems~\cite{casetti}.
In this Rapid Communication, we derive an approximate
analytical expression for $\lambda_1$ 
which agrees very well with numerical simulations.
We focus our 
attention on the asymptotic state of the
system, when  it has reached energy equipartition,
i.e. the state where energy is evenly distributed among all Fourier modes
(relaxation times~\cite{kantz} will not be our concern here).
One of the main points of this paper is to emphasize
the relevant  role played by some unstable periodic orbits
corresponding to Fourier modes.
Therefore, we will first derive the criterion for modulational instability 
of a plane wave on the lattice.

Denoting by $u_n(t)$ the position of the $n$th
atom ($ n \in [1,N]$), the equations of motion of the FPU chain read
\begin{eqnarray}
\nonumber
\ddot{u}_n &&= u_{n+1} + u_{n-1} - 2u_n + \\
&&\beta\left[(u_{n+1}-u_n)^{2p+1} - 
(u_n - u_{n-1})^{2p+1}\right]
\label{sub}
\end{eqnarray} 
where $p$ is an integer greater or equal than 1.
We chose periodic boundary conditions.
Even if the positive parameter $\beta$ can be forgotten by  appropriate 
scaling transformations of $u_n$,
we will keep it in order to make reliable 
comparisons with previous papers, where
$\beta=0.1$. For sake of simplicity, we consider first the case $p=1$ 
and then we generalize to any $p$-value.

Looking for plane wave solutions
\begin{equation}
u_n(t) =\phi_0\left(e^{i\theta_n (t)}+e^{-i\theta_n (t)}\right)
\label{plane}
\end{equation}
where $\theta_n(t) = qn-\omega t$ and $q=2\pi k/N$,
we obtain the dispersion relation
$\omega^2(q)=4(1+\alpha)\sin ^{2} (q/2)$
where $\alpha=12\beta\phi_0^2 \sin^2(q/2)$
takes into account the nonlinearity \cite{pietro}.
The modulational instability of such a plane wave 
is investigated by studying the
linearized equation associated to the
envelope of the carrier wave.
Therefore, one introduces an infinitesimal
perturbation in the amplitude and looks for solutions
\begin{equation} 
u_n(t) = [ \phi_0 + b_n (t) ]\ e^{i\theta_n (t)}+
[ \phi_0 + b_n^\star (t) ]\ e^{-i\theta_n (t)}\quad .
\end{equation}
Introducing this ansatz in Eq.~(\ref{sub}), 
after linearization with respect to $b_n$ but
keeping the second derivative (contrary to what has been
 done for Klein-Gordon type
equation\cite{daumont}),
we obtain 
\begin{eqnarray}
\nonumber
\ddot b_n-2i\omega \dot b_n &&=
(1+2\alpha)(b_{n+1}e^{iq}+b_{n-1}e^{-iq}-2\cos(q)\ b_n) \\
 &&+\alpha(b_{n+1}^*+b_{n-1}^*-2\cos(q)\ b_n)
\end{eqnarray}
Assuming further $b_n=A \ e^{i(Qn-\Omega t)}+B \ e^{-i(Qn-\Omega t)}$,
we finally obtain  the following dispersion relation
\begin{eqnarray}
\nonumber
&&\Biggl[(\Omega+\omega)^2-4(1+2\alpha)\sin^2\left({q+Q\over2}\right)\Biggr] 
\times \\
&&\Biggl[(\Omega-\omega)^2-
4(1+2\alpha)\sin^2\left({q-Q\over2}\right)\Biggr]
= 4\alpha^2\left(\cos{Q}-\cos q\right)^2
\label{relatdispercorr}
\end{eqnarray}

This equation has 4 different solutions once $q$
(wavevector of the unperturbed  wave) and $Q$ (wavevector of the perturbation)
are given.
If one of the solutions is complex we have an instability
of one of the modes $(q\pm Q)$ 
with a growth rate equal to the imaginary part of the solution. Therefore,
one can compute the instability threshold for any initial
linear wave, i.e. any wavevector and any amplitude.
For example for $q=0$ we find that the solution is obviously stable
since the zero-mode, corresponding to translation invariance,
is completely decoupled from the others. For $q=\pi$, the
expression for the growth rate is
\begin{eqnarray}
\nonumber 
\tau(\pi,Q) &&=
2\Bigl(\sqrt{(1+\alpha)(4+8\alpha)\cos^2(Q/2)
+\alpha^2\cos^4(Q/2)} \\
&&-1-\alpha-(1+2\alpha)\cos^2(Q/2)\Bigr)^{1/2}
\label{growthrate}
\end{eqnarray}
A simple analysis of this function shows that the first unstable mode
is the nearest mode corresponding to $Q=2\pi/N$. Computing the critical 
value of the parameter $\alpha$ above which $\tau(\pi,2\pi/N)$ is positive, 
we obtain the critical energy $E_c$ for the $\pi$-mode. It reads 
\begin{equation}
E_c={2N\over9\beta}\sin^2({\pi\over N}){7\cos^2({\pi\over N})-1\over 
3\cos^2({\pi\over N})-1}\quad .
\end{equation}
This analytical expression is in agreement with the previous 
\cite{berman,flach,poggi-ruffo} approximate expression $E_c\simeq{\pi^2/3N\beta}$ 
valid only in the large $N$ limit. 
Above this energy threshold, the $\pi$-mode is therefore unstable
and gives rise to a chaotic localized 
breather-like excitation, able to move very fast
in the system, collecting energy from high-wavevectors
phonons until its disappearance leads to the final 
energy equipartition\cite{cretegny}.

\begin{figure}
\psfig{figure=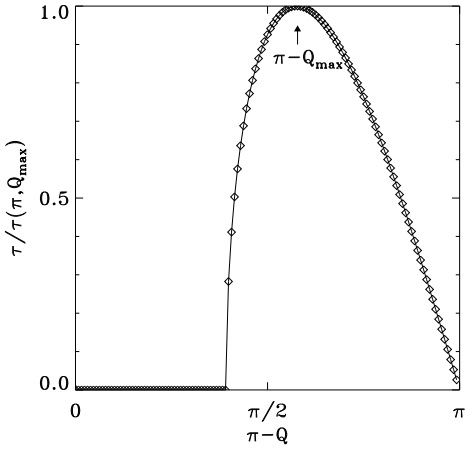,height=6truecm,width=6truecm}
\vskip 0.5truecm
\caption{Shape of the growth rate $\tau(\pi,Q)/\tau(\pi,Q_{\hbox{max}})$
for sufficiently high energy.
The diamonds  corresponds to the case $N=256$ and the solid curve to the
asymptotic shape when $N$ goes to infinity.}
\label{frozenshape}
\end{figure}

When the energy increases, the region of
instability extends to a larger region of wavevectors, and
in particular the most unstable mode $Q_{\hbox{max}}$
increases until the asymptotic value
$Q_{\hbox{max}}=2 \arccos\bigl(\sqrt{8/\sqrt{3}-4}\bigr)\simeq 0.42 \pi$
is reached.  It is important  to notice that,
for sufficiently high energy, the rescaled growth rate
$\tau/\tau(\pi,Q{\hbox{max}})$ does not depend on the energy density. 
The growth rate is plotted in Fig.~\ref{frozenshape} at high energy.

\begin{figure}
\psfig{figure=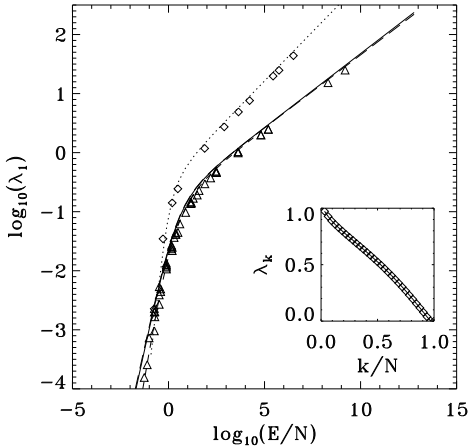,height=6truecm,width=6truecm}
\vskip 0.5truecm
\caption{Comparison of the analytical estimate  with numerical results
for the maximal Lyapunov exponent.
The solid curve corresponds to our estimation 
(Eq.~{\protect \ref{analexpr}}), the dashed
curve to the estimate using Riemannian differential geometry
(see Ref.~{\protect \onlinecite{casetti}})
and the triangles to our numerical results (for the $\beta$-FPU, i.e. $p=1$).
The dotted curve (resp. diamonds)
corresponds to the analytical estimate (resp. numerical results) in
the case of a power $p=2$.
In the inset, we plot the $N$ positive Lyapunov in the case $E/N=4200$ and 
$p=1$.}
\label{numtheo}
\end{figure}

We performed some simulations
of the system with a 6th-order symplectic integration scheme
adopting as initial condition the $\pi$-mode and
computing the Lyapunov exponents after the transition to equipartition.
We used the algorithm proposed by Benettin et al\cite{callyap},
where the full set of tangent vectors is periodically reorthonormalized using
the Gram-Schmidt method; the Lyapunov exponents are then obtained from the
time average of the logarithms of the normalization factors.
The results are  plotted in Fig.~\ref{numtheo}. We have
also checked our results by performing the numerical integration
directly in Fourier space, paying particular attention to the
Lyapunov eigenvectors.

Let us present now the analytical estimation of the maximal Lyapunov.
As the system is symplectic, the usual pairing rule is valid and moreover
Pesin's theorem allows us to identify the
Kolmogorov-Sinai entropy of the system with the sum
of all positive Lyapunov exponents. 
As the spectrum was shown~\cite{LPR} to be approximately linear at high energy
(see the inset of Fig.~\ref{numtheo}),
one can relate the Kolmogorov-Sinai entropy $S_{KS}$
with the maximal Lyapunov exponent, namely
\begin{equation}
S_{KS}=\sum_{i=1}^N\lambda_i\cong{\lambda_1N/2} \quad .
\end{equation}
Let us define the instability entropy
\begin{equation}
S_{IE}(q)=\sum_{i=1}^{N/2}\tau(q,2\pi i/N)\quad,
\end{equation}
where the sum is over all positive growth rates~\cite{posgrowth}.
The crucial physical hypothesis of this paper 
is that $S_{KS} \simeq S_{IE}(\pi)$,
we then obtain the following analytical expression for the
maximal Lyapunov exponent:
\begin{equation}
\lambda_1={2\over N}\sum_{i=1}^{N/2} \tau(\pi,2\pi i/N) \label{analexpr}\quad .
\end{equation}
Using the expression~(\ref{growthrate}), we can then compute
the maximal Lyapunov exponent. 
Fig.~\ref{numtheo} attests that the analytical expression~(\ref{analexpr})
is very accurate. In the same figure the data obtained with a 
completely different approach, developed by Casetti, Livi and
Pettini (CLP)~\cite{casetti}, are also shown. 
The two methods give almost identical results, apart at
very low energy, where the CLP-findings~\cite{casetti} is 
in better agreement with our numerical data. However, 
our approach is definetely simpler and relies on the analysis 
of unstable periodic orbits, while the CLP-one on Riemannian 
differential geometry.

It is remarkable to note that  Chirikov~\cite{chirikov}
found similarly the maximal Lyapunov exponent
of the standard map at high energy
by averaging over the phase space the maximal eigenvalue
associated to the main hyperbolic point. It corresponds
in our case to averaging the growth rate~(\ref{growthrate})
for the unstable periodic orbit $q=\pi$ over the equilibrium 
equipartition state (where all modes have the same weight).
A similar approach is known as Toda criterion~\cite{benettingal}
and although it cannot be used as a signature of chaos, it can 
give an approximate estimation of $\lambda_1$.

In fact, one can understand this average in a better
way by recalling that the modes $\{\pi/2\},\{2\pi/3\},\{\pi\}$
correspond to the simplest unstable periodic
orbits and 
are also the only three one-mode solutions of the $\beta$-FPU
problem. The calculation of the 
instability entropies of this three modes shows that they are
extremely close one to another contrary to 
the value for other modes.
A correct approach would be to apply the zeta-function
formalism~\cite{ruelle} to this system, if feasible.

At high energy,
expression~(\ref{growthrate}) can be simplified
as $\tau(\pi,Q)\simeq\sqrt\alpha \ f(Q)$ where $f(Q)$
is energy independent. Therefore the growth rate scales with
the amplitude $\phi_0$, and as 
$E=N\left(8\phi_0^2+64\beta \phi_0^4\right)$, it means
that the growth rate and therefore the maximal Lyapunov 
$\lambda_1$ scales with $(E/N)^{1/4}$ at high energy.
This result is in contradiction with Ref.~\onlinecite{pettini}
but in agreement with Ref.~\onlinecite{casetti}.

In fact, a similar approach gives also very good results for other
powers $2(p+1)$ in the coupling potential.
The expression of the growth rate is then the same if we use 
$\alpha=\beta {(2p+1)!\over p! (p+1)!}
\bigl(2\phi_0\sin(q/2)\bigr)^{2p}$.
Fig.~\ref{numtheo} shows that the results are once more
in very good agreement with numerical estimates.
One derives easily that the maximal Lyapunov
scales at high energy like
$\lambda_1 \sim (E/N)^{{1\over 2}-{1\over 2(p+1)}}$ \cite{note}.
It is important to stress that in the limit of 
hard potential ($p\rightarrow\infty$) we find
the exponent $1/2$, analogously to billiards\cite{benettin}.
In the low energy limit, however, all models have the same 
scaling behaviors as expected. 

Plotting  $\lambda_1 (E/N)$ in a log-log scale we observe 
(see Fig.~\ref{numtheo}) that
the two asymptotic linear behaviors are separated by a
knee at intermediate energy density. An estimation of this 
transition region can be derived assuming that the
linear and nonlinear contributions to $\omega(q)$ should
be of the same order. We obtain $\alpha\sim1$ i.e.,
an energy density of the order of $1/\beta$
(this is equivalent to the estimation given by mode overlap 
criterion\cite{izrailev}).

\begin{figure}
\psfig{figure=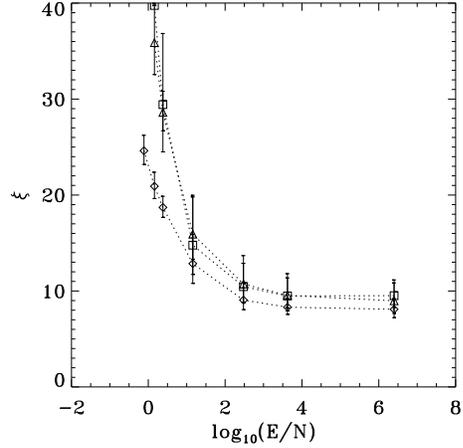,height=6truecm,width=6truecm}
\vskip 0.5truecm
\caption{
Participation ratio $\xi$ 
versus the density of Energy $E/N$, for different lattice sizes;
diamonds for $N=64$, triangles for $N=256$ and squares for
$N=1024$.}
\label{localization}
\end{figure}

\begin{figure}
\psfig{figure=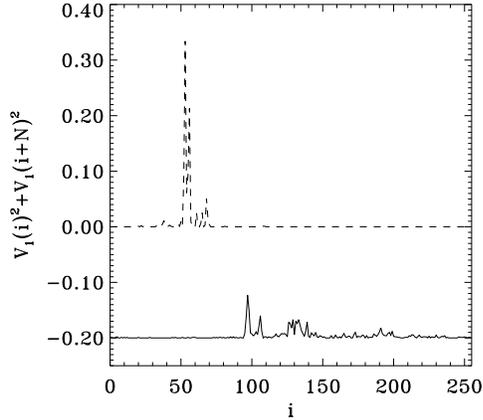,height=6truecm,width=6truecm}
\vskip 0.5truecm
\caption{Localization in tangent space of the Lyapunov
vector for $N=256$.
The dashed curve corresponds to a generic localized Lyapunov
vector at high energy density ${E/N}\simeq 2.5 \ 10^6$,
whereas the solid curve (shifted by -0.2)
corresponds to a generic delocalized one
at low energy density ${E/N}\simeq2$.}
\label{lyapvector}
\end{figure}

This knee corresponds to a stochasticity 
threshold~\cite{livi} which defines the crossing from weak to 
strong chaos.  But we have also found that it corresponds to 
an interesting transition in tangent space. 
Considering the normalized Lyapunov vector $V_1$ associated to the maximal
Lyapunov $\lambda_1$, we can introduce the participation
ratio\cite{prisraielev} 
\begin{equation}
\xi=\left(\sum_{i=1}^N\left[V_1(i)^2+
V_1(i+N)^2\right]^2\right)^{-1}\quad.
\label{lyapvec}
\end{equation}
where the first (resp. last) $N$ components of $V_1$
are associated to the evolution of linear perturbation of
$u_n$ (resp. ${\dot u}_n$) in tangent space.
The quantity (\ref{lyapvec}) has been used in different contexts and
for example in dynamical systems~\cite{kaneko} as
an indicator of localization: it is of order $N$ 
if the vector is extended and of order one if localized.
We have found that the stochasticity threshold
corresponds to a crossover from an extended state
in tangent space to a more localized state, as attested by
Fig.~\ref{localization}. The two examples of Lyapunov vectors
reported in Fig.~\ref{lyapvector} show the clear difference
between low and high energy density.
This transition is very reminiscent of the metallic-insulating 
transition in finite samples\cite{kramer}.

We conclude by stressing again that 
we have  computed the maximal Lyapunov for a high dimensional Hamiltonian
system by using a simple analytical approach, based upon modulation
instability analysis of linear waves. The results obtained here are 
in excellent agreement with our computer simulation results. 
The success of this calculation suggests that this Lyapunov estimation 
could be extended to other high-dimensional Hamiltonian systems. 
Moreover, we have shown that the strong
stochasticity threshold \cite{livi} is {\it not} a threshold to 
energy equipartition, since equipartition can be always obtained,
although on longer time scales \cite{deluca}. It rather corresponds 
to a crossover from extended to more localized state in tangent space.

We would like to thank G. Benettin, P. Poggi and A. Politi for enlightening
discussions as well as L. Casetti, R. Livi and M. Pettini, 
for providing us some of their data.  T.D. gratefully acknowledges 
EC for financial support with grant ERBFMBI-CT96-1063. 
S.R. thanks CICC, Cuernavaca, Mexico and ESI Institute, Vienna,
Austria for hospitality and financial support.
This work is also part of the EC network on ``Stability and 
universality in classical mechanics'' (contract ERBCHRX-CT94-0460).
Part of CPU time has been nicely supplied by the 
Institute of Scientific Interchanges (ISI) of Torino.

\begin{references}
\bibitem{pietro} For $q = \pi/2$, $2 \pi /3$ and $\pi$, solution
(\protect{\ref{plane}}) is the first term of the
harmonic expansion of an exact solution of the $\beta$-FPU
model \cite{poggi-ruffo}. In general an exact
solution can only be recovered by combination of several $q$-modes.
\bibitem{posgrowth} For a given wavevector, there are 4 solutions of 
Eq.~(\ref{relatdispercorr}). Two are real and two complex conjugates:
 i.e. 2 solutions
with positive (or null) growth rates and two with negative (or null)
growth rates.
\bibitem{note} The correct scaling law at high density of energy 
can be also obtained by assuming that the evolution in the tangent
space is ruled by the average value of the Hessian matrix.
A similar mean-field approach was previously
used for billiards~{\protect \cite{benettin}}.
\end{references}
\end{document}